\documentclass[aps,showpacs,preprintnumbers]{revtex4}
\usepackage{amssymb}
\usepackage{amsmath}
\usepackage{bm}

\input{tcilatex}

\begin{document}

\title{\textbf{Why pair production cures covariance in the light-front?}}
\author{J.H.O. Sales}
\altaffiliation{Funda\c{c}\~{a}o de Ensino e Pesquisa de Itajub\'{a}, cep: 37501-002, Itajub%
\'{a}-MG-Brazil.}
\affiliation{Instituto de Ci\^{e}ncias, Universidade Federal de Itajub\'{a}, CEP
37500-000, Itajub\'{a}, MG, Brazil}
\author{A.T. Suzuki}
\affiliation{Department of Physics, North Carolina State University, Raleigh, NC}
\date{\today }

\begin{abstract}
We show that the light-front vaccum is not trivial, and the Fock space for
positive energy quanta solutions is not complete. As an example of this non
triviality we have calculated the electromagnetic current for scalar bosons
in the background field method were the covariance is restored through
considering the complete Fock space of solutions. We also show thus that the
method of ''dislocating the integration pole'' is nothing more than a
particular case of this, so that such an ''ad hoc'' prescription can be
dispensed altogether if we deal with the whole Fock space.

In this work we construct the electromagnetic current operator for a system
composed of two free bosons. The technique employed to deduce these
operators is through the definition of global propagators in the light front
when a background electromagnetic field acts on one of the particles.
\end{abstract}

\pacs{12.39.Ki,12.20.-m,21.45.+v}
\maketitle

\address{ 27607.}

%\preprint{APS/123-QED}






%\twocolumn
%\narrowtext

\section{Introduction}

In the traditional approach to restore covariance of the electromagnetic
current in the light front an \textquotedblright ad hoc\textquotedblright\
prescription of dislocating the pole is employed \cite{9}. However, this
procedure of \textquotedblright pole dislocation\textquotedblright\ has no
physical grounds and the arrival at the correct result is just fortuitous.
We demonstrate that the light front Fock space of positive quanta solutions
is incomplete and that as a consequence the non triviality of the light
front vacuum is a mandatory feature in the new scenario.

In order to do this we calculate in a specific example the matrix element
for the electromagnetic current in the Breit's reference frame for $%
q^{+}\rightarrow 0$ and $\vec{q}_{\bot }\neq 0$. To this end we use a
constant vertex for the bound state of two bosons in the light front. Such
calculation agrees with the results obtained through the computation of the
triangular diagram for the electromagnetic current of a composite boson
whose vertex is constant \cite{9}.

Sawicki \cite{37} has shown that in the Breit's reference frame, the $%
J^{+}=J^{0}+J^{3}$ component of the electromagnetic current for the bound
state of two bosons, obtained from the triangular Feynman diagram after
integration in the $k^-$ component of the loop momentum, has no pair
production contribution from the photon. As a consequence of this fact, the
electromagnetic form factor, calculated in the light front, starting from $%
J^+$ is identical to the one obtained in the covariant calculation. By
covariant calculation of an amplitude we mean the computation of momentum
loop integrated directly without the use of the transformation into light
front momentum.

The problem that appears when integrating in the light front coordinates in
momentum loops was studied by Chang and Yan \cite{33} and more recently
discussed in references \cite{34,35,9,51}. In the Chang and Yan's works
although it is pointed out the difficulty in the $k^{-}$ integration for
certain amplitudes and suggested a possible solution to the problem, in our
view two distinct aspects are mingled together which are the renormalization
question and the problem of integration in the light front coordinates. Our
emphasis here is the covariance restoration for the electromagnetic current
through a careful integration in the $k^{-}$ of the loop momentum in finite
diagrams. We know that for the $J^{+}$ component of the electromagnetic
current for a particle of spin $1$ there are terms that correspond to pair
production in the light front formalism for $q^{+}=0$ \cite{9,51}. In the
case of vector meson $\rho $, the rotational invariance of the current $%
J^{+} $ is broken when we use the light front formalism unless pair
production diagrams are duly considered \cite{46a,41}.

In references \cite{37,38,40,39} the electromagnetic current in the light
front for a composite system is obtained from the triangular diagram
(impulse approximation) when this is integrated in the internal loop
momentum component $k^{-}(=k^{0}-k^{3})$. This integration in $k^{-}$ by
Cauchy's theorem uses the pole of the spectator particle in the process of
photon absorption for $q^{+}=0$. Using the current $J^{+}$ the process of
pair creation by the photon is in principle eliminated \cite{37,38}. In
general, covariance is preserved under kinematic transformations, but the
current looses this physical property under a more general transformation
such as rotations and parity transformations. We show how the pair
production is necessary to the complete calculation of the current's $J^{-}$
component in the Drell-Yan's reference frame ($q^{+}=0$).

In section $2$ we describe the electromagnetic current of two
non-interacting bosons and also derive the diagrams for pair production. In
section $3$, we construct the electromagnetic current for a bound state with
constant vertex in the limit of photon's momentum transfer $q^{+}\rightarrow
0$, where $q^{+}=q^{0}+q^{3}$. To this purpose we use the Breit's reference
frame. In that limit, the pair production survives for the $j^{-}$ current
and accounts for the covariance of the current in this model.

\section{Generating functional for interacting fields}

Long ago Dirac \cite{dirac} has shown that dynamical forms can be built
using the hyperplane called null plane defined by $x^{+}=t+z=0$ which is the
\textquotedblright time\textquotedblright\ coordinate for the light front
instead of the usual hypersurface $t=0$. In that light front form, a boost
in the $z$-direction does not modify the null plane.

In this framework we construct the electromagnetic current operator for the
system composed of two free bosons in the light front. The technique we use
to deduce such operators is to define the global propagators in the light
front when a electromagnetic background field acts on one of the particles.
Although we are in fact calculating the global propagator for two bosons in
an electromagnetic background field, we extrapolate the language using terms
such as \emph{\textquotedblleft current operator\textquotedblright } and 
\emph{\textquotedblleft current\textquotedblright } to designate such
operations. We show that for the $J^{-}$ case the two free boson propagators
in a background field have a contribution from the process of photon's pair
production, being crucial to restore current's covariance.

The normalized generating functional is given by 
\begin{equation}
Z\left[ J\right] =\frac{\int \mathcal{D}\phi \exp \left[ i\mathcal{S}+i\int
J\phi dx \right] }{\int \mathcal{D}\phi e^{iS}}  \label{3.3.2}
\end{equation}
where $\mathcal{S} =\int \pounds dx$. It is easy to note that when $\pounds %
_{\text{int}}=0$ we have the expression $Z_{0}\left[ J\right] $ for the free
particle (\ref{cov}). So, we can find the propagators, or Green's funstions,
in an electromagnetic field. 
\begin{equation}
S\left( x^{+}\right) =-ieA^{\mu }\int d\overline{x}^{+}\left[ S_{3}\left(
x^{+}-\overline{x}^{+}\right) \frac{\partial S_{1}\left( \overline{x}
^{+}\right) }{\partial \overline{x}^{\mu }}-S_{1}\left( \overline{x}
^{+}\right) \frac{\partial S_{3}\left( x^{+}-\overline{x}^{+}\right) }{
\partial \overline{x}^{\mu }}\right] \times S_{2}\left( x^{+}\right)  \notag
\end{equation}

\subsection{Current $J^{-}$}

The $"-"$ component of the electromagnetic current operator for the system
of two bosons is obtained from the propagator of this system in the
situation in which one of the bosons interacts with a background
electromagnetic field that has only component $"+"$. This process is
represented by the propagation of a composite system from $0$ to $x^{+}$,
without mutual interaction between bosons and the interaction with the
electromagnetic field ocurring at $\overline{x}^{+}$. The electromagnetic
current operator $J^{-}$ for one boson is given by $-ie\left( \frac{%
\overrightarrow{\partial }}{\partial \overline{x}^{+}}-\frac{\overleftarrow{%
\partial }}{\partial \overline{x}^{+}}\right) $. The propagator in the light
front where one of the bosons interacts with the background field $(A^{+})$
is given by 
\begin{eqnarray}
S\left( x^{+}\right) &=&-ie\int dq^{-}F\left( q^{-}\right) \int d\overline{x}
^{+}e^{-\frac{iq^{-}\overline{x}^{+}}{2}}\times  \label{6.1} \\
&&\left[ S_{3}\left( x^{+}-\overline{x}^{+}\right) \frac{\partial
S_{1}\left( \overline{x}^{+}\right) }{\partial \overline{x}^{+}}-S_{1}\left( 
\overline{x}^{+}\right) \frac{\partial S_{3}\left( x^{+}-\overline{x}
^{+}\right) }{\partial \overline{x}^{+}}\right] \times S_{2}\left(
x^{+}\right)  \notag
\end{eqnarray}
where $S_{3}\left( x^{+}-\overline{x}^{+}\right) $ is the boson propagator
after the interaction with the background field and $S_{1}\left( \overline{x}%
^{+}\right) $ before interaction. The function $F(q^{-})$ is the Fourier
transform of the external field that is defined as: 
\begin{equation}
A^{+}=\int dq^{-}F\left( q^{-}\right) e^{-\frac{iq^{-}\overline{x}^{+}}{2}}
\end{equation}

Introducing the free particle propagators $S_{i}$, given by (\ref{lf1}),
with $i=1,2,3$ into equation (\ref{6.1}) and integrating over $\bar{x}^{+}$
there appears a Dirac delta which corresponds to momenta \textquotedblleft $%
-^{\prime \prime }$ conservation in the vertex of the coupling with the
electromagnetic field. With this we reduce the three integrals to two
integrals in $k^{-}$. The remnant integations are in $k_{1}^{-}$ and $%
k_{2}^{-}$ which are the momenta for the initial particles. In the integrand
we have the exponential function $\exp [-i\frac{\left(
k_{1}^{-}+k_{2}^{-}+q^{-}\right) x^{+}}{2}]$ which will be integrated over
when we do the Fourier transform, i.e., we make use of the transform 
\begin{equation}
\widetilde{S}(K^{-})=\int dx^{+}S\left( x^{+}\right) e^{i\frac{K^{-}x^{+}}{2}%
},  \label{6.3}
\end{equation}%
so that, as a result, we have another Dirac's delta, corresponding to the
conservation of total $K^{-}$, that is $K^{-}=K_{i}^{-}+q^{-}$. This is the
law of energy conservation in the light front, where $K_{i}^{-}$ is the
total initial energy and $q^{-}$ the momentum transferred by the photon.
Thus, there is one more integration reduction leaving only the the $%
k_{2}^{-} $ integration, often called momentum of the spectator particle.
This is so because we couple the background field to particle 1, so that the
particle 2 only \textquotedblleft observes\textquotedblright\ this coupling.
From transformation (\ref{6.3}) we have 
\begin{eqnarray}
\widetilde{S}(K^{-}) &=&\frac{ei^{4}}{8\pi }\int dq^{-}F(q^{-})\left[ \int 
\frac{dk_{2}^{-}}{k_{2}^{+}(K_{i}^{+}-k_{2}^{+})(K^{+}-k_{2}^{+})}\times
\right.  \label{6.3a} \\
&&\left. \frac{2(K_{i}^{-}-k_{2}^{-})+q^{-}}{(K_{i}^{-}-k_{2}^{-}-\frac{%
m_{1\bot }^{2}-i\varepsilon }{K_{i}^{+}-k_{2}^{+}})(k^{-}-\frac{m_{2\bot
}^{2}-i\varepsilon }{k_{2}^{+}})(K^{-}-k_{2}^{-}-\frac{m_{3\bot
}^{2}-i\varepsilon }{K^{+}-k_{2}^{+}})}\right] ,  \notag
\end{eqnarray}%
where $K_{i}^{-}=k_{1}^{-}+k_{2}^{-},K^{-}=k_{2}^{-}+k_{3}^{-}$, $%
k_{3}^{-}-k_{1}^{-}=q^{-}$, $m_{j\bot }^{2}=k_{j\bot }^{2}+m^{2}$ and $%
j=1,2,3$.

The component of the current operator $J^{-}$ is obtained from the operator $%
\mathcal{O}^{-}$ which is represented by the braces in (\ref{6.3a}), so that 
\begin{eqnarray}
\mathcal{O}^{-} &=&C_{0}^{\times }\int \frac{dk_{2}^{-}}{
k^{+}(K_{i}^{+}-k_{2}^{+})(K^{+}-k_{2}^{+})}\times  \label{6.5} \\
&&\frac{2(K_{i}^{-}-k_{2}^{-})+q^{-}}{(K_{i}^{-}-k_{2}^{-}-\frac{m_{1\bot
}^{2}-i\varepsilon }{K_{i}^{+}-k_{2}^{+}})(k_{2}^{-}-\frac{m_{2\bot
}^{2}-i\varepsilon }{k_{2}^{+}})(K^{-}-k_{2}^{-}-\frac{m_{3\bot
}^{2}-i\varepsilon }{K^{+}-k_{2}^{+}})},  \notag
\end{eqnarray}
where $C_{0}^{\times }=\frac{(-ie)}{8\pi }$ is a constant that contains
terms resulting from the integration in $\bar{x}^{+}$ and integrations in $%
k_{1}^{-}$ and $k_{3}^{-}$. The index $\times $ signifies an external field
and the lower index $0$ stands for the absence of $\sigma $ exchange bosons.

The integration is performed using the Cauchy theorem in the $k_{2}^{-}$
complex plane in two regions for $k_{2}^{+}$, where in particular we use $%
q^{+}>0$, so that

I) $0<k_{2}^{+}<K_{i}^{+}<K^{+}$ and

II) $0<K_{i}^{+}<k_{2}^{+}<K^{+}$.

The poles are $k_{2}^{-}=\frac{k_{\perp }^{2}+m^{2}}{k_{2}^{+}}$ for the
first region of integration and $k_{2}^{-}=K^{-}-\frac{(K-k_{2})_{\bot
}^{2}+m^{2}}{(K-k_{2})^{+}}$ for the second. Note that for the latter region
we clearly have the demand that $k_{2}^{+}>-k_{1}^{+}$ which means that
naturally the Fock space for positive quanta solutions is incomplete.

For the first region of integration $(I)$ the final result reads: 
\begin{eqnarray}
\mathcal{O}_{I}^{-} &=&\frac{(-ie)\theta (K_{i}^{+}-k_{2}^{+})\theta (k^{+}) 
}{4k^{+}(K_{i}^{+}-k_{2}^{+})(K^{+}-k_{2}^{+})}\times \frac{i}{(K^{-}-\frac{
m_{2\perp }^{2}-i\varepsilon }{k_{2}^{+}}-\frac{m_{3\bot }^{2}-i\varepsilon 
}{K^{+}-k_{2}^{+}})}  \label{j1} \\
&&\times \left[ 2\left( K_{i}^{-}-\frac{m_{2\perp }^{2}}{k_{2}^{+}}\right)
+q^{-}\right] \times  \notag \\
&&\frac{i}{(K_{i}^{-}-\frac{m_{2\perp }^{2}-i\varepsilon }{k_{2}^{+}}-\frac{
m_{1\bot }^{2}-i\varepsilon }{K_{i}^{+}-k_{2}^{+}})},  \notag
\end{eqnarray}
where onde $m_{1\perp }^{2}=\left( K_{i}-k_{2}\right) _{\perp }^{2}+m^{2}$, $%
m_{2\perp }^{2}=k_{2\perp }^{2}+m^{2}$, $m_{3\perp
}^{2}=\left(K-k_{2}\right) _{\perp }^{2}+m^{2}$.

We see that in (\ref{j1}) that there is global propagation of two bodies
forward to future before and after the interaction with the external field.
In the initial state the total energy in the light front is $K_{i}^{-}$.
After the absorption of the photon, the bosons propagate with momenta $%
k_{3}^{-}$, $k_{2}^{-}$ and final total energy $K^{-}=K_{i}^{-}+q^{-}$.

Evaluating (\ref{6.5}) by residue integration for the second region, $(II)$,
we have the other component of the operator $\mathcal{O}^{-}$, given by 
\begin{eqnarray}
\mathcal{O}_{II}^{-} &=&\frac{(-ie)\theta (K^{+}-k_{2}^{+})\theta
(k_{2}^{+}-K_{i}^{+})\theta (k^{+})}{
4k_{2}^{+}(k_{2}^{+}-K_{i}^{+})(K^{+}-k_{2}^{+})}\frac{i}{(K^{-}-\frac{
m_{2\perp }^{2}-i\varepsilon }{k_{2}^{+}}-\frac{m_{3\bot }^{2}-i\varepsilon 
}{K^{+}-k_{2}^{+}})}  \notag \\
&&\times \left[ -2\left( K^{-}-K_{i}^{-}-\frac{m_{3\perp }^{2}}{
K^{+}-k_{2}^{+}}\right) +q^{-}\right] \times  \notag \\
&&\frac{i}{(K^{-}-K_{i}^{-}-\frac{m_{3\bot }^{2}-i\varepsilon }{
K^{+}-k_{2}^{+}}-\frac{m_{1\bot }^{2}-i\varepsilon }{k_{2}^{+}-K_{i}^{+}})}.
\label{j2}
\end{eqnarray}

We observe that the operator (\ref{j2}) has a propagation in the
intermediate state of a particle and an anti-particle given by $%
(K^{-}-K_{i}^{-}-\frac{m_{3\bot }^{2}}{K^{+}-k_{2}^{+}}-\frac{m_{1\bot }^{2}%
}{k_{2}^{+}-K_{i}^{+}})^{-1}$ whose $-$ anti-particle energy is $\frac{%
m_{1\bot }^{2}}{k_{2}^{+}-K_{i}^{+}}$. We remind that for $x^{+}>0$ and $%
k_{1}^{+}>0$, the propagation is forward in time. Otherwise, for $x^{+}<0$
and $k_{1}^{+}<0$ we have the propagation of anti-particles backward in
time. Particle $1$ has light-front energy given by 
\begin{equation}
k_{1}^{-}=\frac{\left( K_{i}-k_{2}^{+}\right) _{\perp }^{2}+m^{2}}{
K_{i}^{+}-k_{2}^{+}}=\frac{m_{1\bot }^{2}}{K_{i}^{+}-k_{2}^{+}},  \label{p1}
\end{equation}
whereas in the case of the second region of integration we have $%
K_{i}^{+}-k_{2}^{+}<0$ which leads us to $k_{1}^{-}<0$, that is, we have a
propagation to the past of an particle in a time on the light front $x^{+}<0$%
, corresponding to physical anti-particle propagating forward in time.

We can interpret this result as a pair production in a light front time
previous to $x^{+}=0$ which is the initial time define in the null plane.
Figure (\ref{fig1a}) depicts the diagram corresponding to the operator $%
\mathcal{O}_{I}^{-}$ and figure (\ref{fig2a}) the diagram corresponding to
the propagator $\mathcal{O}_{II}^{-}$. We see in this figure that in a time
before $x^{+}=0$ we have pair production due to the photon. Later on we will
discuss the importance of equation (\ref{j2}) for the covariance of the
electromagnetic current in the Breit's reference frame.

\FRAME{ftbpFU}{2.7223in}{0.8908in}{0pt}{\Qcb{Propagation of two bosons in
background field.}}{\Qlb{fig1a}}{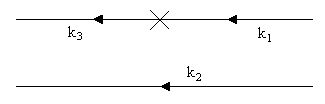}{\special{language "Scientific
Word";type "GRAPHIC";maintain-aspect-ratio TRUE;display "USEDEF";valid_file
"F";width 2.7223in;height 0.8908in;depth 0pt;original-width
2.6002in;original-height 0.8319in;cropleft "0";croptop "1";cropright
"1";cropbottom "0";filename 'dia1.jpg';file-properties "XNPEU";}}

\FRAME{ftbpFU}{2.6318in}{1.0652in}{0pt}{\Qcb{Pair production in a light
front time previous to $x^{+}=0$.}}{\Qlb{fig2a}}{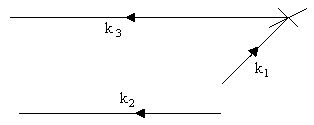}{\special{language
"Scientific Word";type "GRAPHIC";maintain-aspect-ratio TRUE;display
"USEDEF";valid_file "F";width 2.6318in;height 1.0652in;depth
0pt;original-width 2.5114in;original-height 1.0004in;cropleft "0";croptop
"1";cropright "1";cropbottom "0";filename 'dia2.jpg';file-properties
"XNPEU";}}

\subsection{Current $J^{+}$}

The $J^{+}$ component of the electromagnetic current for two bosons
propagating to the future in the light front is calculated in a similar
manner as done in the previous section, with the difference that the
coupling now is $-ie\left( \frac{\overrightarrow{\partial }}{\partial 
\overline{x}^{-}}-\frac{\overleftarrow{\partial }}{\partial \overline{x}^{-}}%
\right) $, so that the propagator for two bosons in the background field is
given by 
\begin{eqnarray}
S\left( x^{+}\right) &=&-ie\int dq^{-}F\left( q^{-}\right) \int d\overline{x}
^{+}e^{-\frac{iq^{-}\overline{x}^{+}}{2}}\times  \label{6.9} \\
&&\left[ S_{3}\left( x^{+}-\overline{x}^{+}\right) \frac{\partial
S_{1}\left( \overline{x}^{+}\right) }{\partial \overline{x}^{-}}-S_{1}\left( 
\overline{x}^{+}\right) \frac{\partial S_{3}\left( x^{+}-\overline{x}
^{+}\right) }{\partial \overline{x}^{-}}\right] \times S_{2}\left(
x^{+}\right) .  \notag
\end{eqnarray}

Integrating in $\overline{x}^{+}$ and doing the Fourier transform we have 
\begin{eqnarray}
\widetilde{S}(K^{-}) &=&C_{0}^{\times }\int dq^{-}F(q^{-})\left[ \frac{
dk_{2}^{-}}{k_{2}^{+}(K_{i}^{+}-k_{2}^{+})(K^{+}-k_{2}^{+})}\times \right. \\
&&\left. \frac{2(K_{i}^{+}-k_{2}^{+})+q^{+}}{(K_{i}^{-}-k_{2}^{+}-\frac{
m_{1\bot }^{2}-i\varepsilon }{K_{i}^{+}-k_{2}^{+}})(k_{2}^{-}-\frac{m_{2\bot
}^{2}-i\varepsilon }{k_{2}^{+}})(K^{-}-k_{2}^{-}-\frac{m_{3\bot
}^{2}-i\varepsilon }{K^{+}-k_{2}^{+}})}\right] .  \notag
\end{eqnarray}

The procedure to calculate the integration via Cauchy's theorem in $%
k_{2}^{-} $ is the same as the one used for the current $J^{-}$. Therefore
we have two components for the current $J^{+}$, one for each region of the
complex $k_{2}^{-}$ plane with the pole positions defined by the intervals $%
0<k_{2}^{+}<K_{i}^{+}<K^{+}$ and $0<K_{i}^{+}<k_{2}^{+}<K^{+}$. The result
is 
\begin{eqnarray}
\mathcal{O}_{I}^{+} &=&\frac{(-ie)\theta (K_{i}^{+}-k_{2}^{+})\theta
(k_{2}^{+})}{4k_{2}^{+}(K_{i}^{+}-k_{2}^{+})(K^{+}-k_{2}^{+})}\frac{i}{
(K^{-}-\frac{m_{2\perp }^{2}-i\varepsilon }{k_{2}^{+}}-\frac{m_{3\bot
}^{2}-i\varepsilon }{K^{+}-k_{2}^{+}})}  \label{6.10} \\
&&\times \left[ 2\left( K_{i}^{+}-k_{2}^{+}\right) +q^{+}\right] \times 
\notag \\
&&\frac{i}{(K_{i}^{-}-\frac{m_{2\perp }^{2}-i\varepsilon }{k_{2}^{+}}-\frac{
m_{1\bot }^{2}-i\varepsilon }{K_{i}^{+}-k_{2}^{+}})},  \notag
\end{eqnarray}
and 
\begin{eqnarray}
\mathcal{O}_{II}^{+} &=&\frac{(-ie)\theta (K^{+}-k_{2}^{+})\theta
(k_{2}^{+}-K_{i}^{+})}{4k_{2}^{+}(k_{2}^{+}-K_{i}^{+})(K^{+}-k_{2}^{+})} 
\frac{i}{(K^{-}-\frac{m_{2\perp }^{2}-i\varepsilon }{k_{2}^{+}}-\frac{
m_{3\bot }^{2}-i\varepsilon }{K^{+}-k_{2}^{+}})}  \label{6.11} \\
&&\times \left[ 2\left( K_{i}^{+}-k_{2}^{+}\right) +q^{+}\right] \times 
\notag \\
&&\frac{i}{(K^{-}-K_{i}^{-}-\frac{m_{3\bot }^{2}-i\varepsilon }{
K^{+}-k_{2}^{+}}-\frac{m_{1\bot }^{2}-i\varepsilon }{k_{2}^{+}-K_{i}^{+}})}.
\notag
\end{eqnarray}

The difference between the operators $\mathcal{O}_{I}^{-}$, $\mathcal{O}%
_{II}^{-}$ and $\mathcal{O}_{I}^{+}$, $\mathcal{O}_{II}^{+}$ is in the
numerators of (\ref{6.10}) and (\ref{6.11}) which have components ''$+$'' in
place of the components ''$-"$. Equation (\ref{6.10}) has only propagation
of particles, both before and after the interaction with the external field.
In equation (\ref{6.11}), however, there is propagation of a particle and an
anti-particle created by the photon of the external field, before the time $%
x^{+}=0$, as it happens with the operator $\mathcal{O}_{II}^{-}$.

\subsection{Current $J_{\perp }$}

The transverse component of the electromagnetic current for the bound state
of two bosons is obtained in an analogous form to the one leading to (\ref%
{6.1}) where the derivative coupling is given by $-ie\left( \frac{ 
\overrightarrow{\partial }}{\partial \overline{x}_{\bot }}-\frac{%
\overleftarrow{\partial }}{\partial \overline{x}_{\bot }}\right) $. The
propagator $S(x^{+})$ of the system with two bosons in a background
electromagnetic field, with only transverse components is given by: 
\begin{eqnarray}
S\left( x^{+}\right) &=&-ie\int dq^{-}F\left( q^{-}\right) \int d\overline{x}
^{+}e^{-\frac{iq^{-}\overline{x}^{+}}{2}}\times \\
&&\left[ S_{3}\left( x^{+}-\overline{x}^{+}\right) \frac{\partial
S_{1}\left( \overline{x}^{+}\right) }{\partial \overline{x}_{\bot }}
-S_{1}\left( \overline{x}^{+}\right) \frac{\partial S_{3}\left( x^{+}- 
\overline{x}^{+}\right) }{\partial \overline{x}_{\bot }}\right] \times
S_{2}\left( x^{+}\right) .  \notag
\end{eqnarray}

Integrating in $\overline{x}^{+}$ and performing Fourier transform, we have 
\begin{eqnarray}
\widetilde{S}(K^{-}) &=&C_{0}^{\times }\int dq^{-}F(q^{-})\left[ \frac{
dk_{2}^{-}}{k_{2}^{+}(K_{i}^{+}-k_{2}^{+})(K^{+}-k_{2}^{+})}\times \right. \\
&&\left. \frac{2(K_{\bot }-k_{2\perp })+q_{\perp }}{(K_{i}^{-}-k_{2}^{-}- 
\frac{m_{1\bot }^{2}-i\varepsilon }{K_{i}^{+}-k_{2}^{+}})(k_{2}^{-}-\frac{
m_{2\bot }^{2}-i\varepsilon }{k_{2}^{+}})(K^{-}-k_{2}^{-}-\frac{m_{3\bot
}^{2}-i\varepsilon }{K^{+}-k_{2}^{+}})}\right] .  \notag
\end{eqnarray}

The integration regions in $k_{2}^{+}$ which determine the positions of the
poles in $k_{2}^{-}$ sre $I)$ $0<k_{2}^{+}<K_{i}^{+}<K^{+}$ and $II)$ $%
0<K_{i}^{+}<k_{2}^{+}<K^{+}$. The first interval corresponds to the
spectator particle be on mass shell and the pole that contributes to the
integration in $k^{-}$ is given by 
\begin{equation*}
k_{2}^{-}=\frac{k_{\perp }^{2}+m^{2}-i\varepsilon }{k_{2}^{+}},
\end{equation*}
so that for this first interval in $k_{2}^{+}$, we have the operator 
\begin{eqnarray}
\mathcal{O}_{I}^{\bot } &=&\frac{(-ie)\theta (K_{i}^{+}-k_{2}^{+})\theta
(k_{2}^{+})}{4k_{2}^{+}(K_{i}^{+}-k_{2}^{+})(K^{+}-k_{2}^{+})}\frac{i}{
(K^{-}-\frac{m_{2\perp }^{2}-i\varepsilon }{k_{2}^{+}}-\frac{m_{3\bot
}^{2}-i\varepsilon }{K^{+}-k_{2}^{+}})} \\
&&\times \left[ 2\left( K_{i}-k_{2}^{+}\right) _{\bot }+q_{\bot }\right]
\times  \notag \\
&&\frac{i}{(K_{i}^{-}-\frac{m_{2\perp }^{2}-i\varepsilon }{k_{2}^{+}}-\frac{
m_{1\bot }^{2}-i\varepsilon }{K_{i}^{+}-k_{2}^{+}})}.  \notag
\end{eqnarray}

Integrating for the second interval we have 
\begin{eqnarray}
\mathcal{O}_{II}^{\bot } &=&\frac{(-ie)\theta (K^{+}-k_{2}^{+})\theta
(k_{2}^{+}-K_{i}^{+})}{4k^{+}(k_{2}^{+}-K_{i}^{+})(K^{+}-k_{2}^{+})}\frac{i}{
(K^{-}-\frac{m_{2\perp }^{2}-i\varepsilon }{k_{2}^{+}}-\frac{m_{3\bot
}^{2}-i\varepsilon }{K^{+}-k_{2}^{+}})} \\
&&\times \left[ 2\left( K_{i}-k_{2}^{+}\right) _{\bot }+q_{\bot }\right]
\times  \notag \\
&&\frac{i}{(K^{-}-K_{i}^{-}-\frac{m_{3\bot }^{2}-i\varepsilon }{
K^{+}-k_{2}^{+}}-\frac{m_{1\bot }^{2}-i\varepsilon }{k_{2}^{+}-K_{i}^{+}})} 
\notag
\end{eqnarray}
where the residue was calculated for the pole 
\begin{equation*}
k_{2}^{-}=K^{-}-\frac{(K-k_{2})_{\bot }^{2}+m^{2}-i\varepsilon }{
(K-k_{2})^{+}}.
\end{equation*}

The operator $\mathcal{O}_{II}^{\bot }$ contains the propagation of a
particle and an anti-particle in the presence of a background field as
commented upon for the operators $\mathcal{O}_{II}^{-}$ and $\mathcal{O}%
_{II}^{+}$.

\section{Zero mode contribution in order $g^{0}$}

In this section we obtain the same result as the work of Sawicki \cite{37}
for the components $J^{+}$ and $J^{\perp}$ of the electromagnetic current
for the case of bound state of two bosons. These components do not present
problems when performing integration in $k^{-}$, that is, we have the same
result as the covariant calculation. In these components for the momentum
transfer $q^{+}=0$, the pair production term vanishes. This is not the case
for $J^-$, as it shall be demonmstrated.

For the current components $J^{+}$ and $\overrightarrow{J}_{\perp }$, only
the contribution to the residue corresponding to particles propagating
forward in time survive in the $k^{-}$ integration. Therefore for these
components there is no contribution from the pair production in the limit $%
q^{+}\rightarrow 0$.

On the other hand, we show that for the current component $J^{-}$, we have
to include in the computation of the diagram of the propagator of two bosons
in a background field, the pair production term which survives in the limit $%
q^{+}\rightarrow 0$. The $J^{-}$ component is sometimes referred to as the
''bad'' component of the electromagnetic current \cite{47,48,49} just
because in general it has the contribution of pairs. Covariance restoration
in the light front implies that for this component of the current, the pair
creation term must be included in the limit $q^{+}\rightarrow 0$.

The matrix element for the electromagnetic current in the Breit's reference
frame, $q^{+}\rightarrow 0$, with one constant vertex $\Gamma $, is
calculated using the operators $\mathcal{O}^{-}, \mathcal{O}^{+}$ and $%
\mathcal{O}^{\bot }$ in order $g^{0}$. The current components can be
calculated as: 
\begin{equation}
J^{-,+,\bot }=\left\langle \Gamma \left| \mathcal{O}^{-,+,\bot }\right|
\Gamma \right\rangle ,  \label{m0a}
\end{equation}
where $\left| \Gamma \right\rangle $ is the ``ket'' for the vertex.

Introducing in (\ref{m0a}) the unit resolution we have 
\begin{eqnarray}
\left\langle \Gamma \left| \mathcal{O}^{-,+,\bot }\right| \Gamma
\right\rangle &=&\int dk^{+}d^{2}k_{\bot }\left\langle \Gamma \right. \left|
k^{+},\overrightarrow{k}_{\bot }\right\rangle \left\langle k^{+}, 
\overrightarrow{k}_{\bot }\right| \times  \notag \\
&&\mathcal{O}^{-,+,\bot }\int dk^{^{\prime }+}d^{2}k_{\bot }^{\prime }\left|
k^{^{\prime }+},\overrightarrow{k^{\prime }}_{\bot }\right\rangle
\left\langle k^{^{\prime }+},\overrightarrow{k^{\prime }}_{\bot }\right.
\left| \Gamma \right\rangle  \notag \\
&=&\Gamma \int dk^{+}d^{2}k_{\bot }\int dk^{^{\prime }+}d^{2}k_{\bot
}^{\prime }\times  \notag \\
&&\left\langle k^{+},\overrightarrow{k}_{\bot }\right| \mathcal{O}^{-,+,\bot
}\left| k^{^{\prime }+},\overrightarrow{k^{\prime }}_{\bot }\right\rangle
\Gamma  \notag \\
&=&\Gamma ^{2}\int dk^{+}d^{2}k_{\bot }dk^{^{\prime }+}d^{2}k_{\bot
}^{\prime }  \notag \\
&&\delta (k^{+}-k^{^{\prime }+})\delta (\overrightarrow{k}_{\bot }- 
\overrightarrow{k^{\prime }}_{\bot }-\overrightarrow{q}_{\bot })\left\langle
k^{+},\overrightarrow{k}_{\bot }\right| \mathcal{O}^{-,+,\bot }\left|
k^{^{\prime }+},\overrightarrow{k^{\prime }}_{\bot }\right\rangle  \notag \\
&=&\Gamma ^{2}\int dk^{+}d^{2}k_{\bot }\left\langle k^{+},\overrightarrow{k}
_{\bot }\right| \mathcal{O}^{-,+,\bot }\left| k^{+},\overrightarrow{k}_{\bot
}+\overrightarrow{q}_{\bot }\right\rangle  \notag \\
&\equiv &\Gamma ^{2}\int dk^{+}d^{2}k_{\bot }\mathcal{O}^{-,+,\bot }.
\end{eqnarray}

Let us now integrate our propagator in $k^{+}$ for $J^{-}$ in the Breit's
reference frame, $q^{-}=q^{+}\rightarrow 0$ and $\overrightarrow{q}_{\bot}=%
\overrightarrow{K}_{\bot }-\overrightarrow{K}_{i\bot }$, then 
\begin{equation}
J^{-}=\Gamma ^{2}\int d^{2}k_{\bot }dk^{+}(\mathcal{O}_{I}^{-}+\mathcal{O}
_{II}^{-}).  \label{jm}
\end{equation}

The operators $\mathcal{O}_{I}^{-}$ and $\mathcal{O}_{II}^{-}$ are given by (%
\ref{j1}) and (\ref{j2}), respectively. We remind ourselves that only the
second operator has pair production.

Changing variables, $x=\frac{k^{+}-K_{i}^{+}}{q^{+}},$ in the second term of
(\ref{jm}), we have 
\begin{equation*}
J_{II}^{-}=\Gamma ^{2}\int d^{2}k_{\bot }\frac{1}{K_{i}^{+}}\int_{0}^{1} 
\frac{dx\theta (x)\theta (1-x)}{(1-x)m_{1\bot }^{2}+xm_{3\bot }^{2}},
\end{equation*}
where $m_{1\perp }^{2}=(K_{i}-k)_{\perp }^{2}+m^{2}$ and $%
m_{3\perp}^{2}=(K-k)_{\perp }^{2}+m^{2}$. Integrating in $x$, we have 
\begin{equation}
J_{II}^{-}=-2\pi \frac{1}{K_{i}^{+}}\int d^{2}k_{\bot }\frac{\ln \left| 
\frac{m_{3\bot }^{2}}{m_{1\bot }^{2}}\right| }{m_{3\bot }^{2}-m_{1\bot }^{2}}%
.
\end{equation}

Since $J_{II}^{-}$ is different from zero, it means that physically in this
component of the electromagnetic current the pair creation is not supressed
in the limit $q^{+}\rightarrow 0$. This result, obtained through the
propagator of two bosons in a background field agrees with our previous work 
\cite{9}. In that work we have used the triangular diagram directly.

In the case of $J^{+}$ we have that the contribution of the pair production
in the limit $q^{+}\rightarrow 0$ is 
\begin{eqnarray}
J_{II}^{+} &=&\frac{\Gamma ^{2}}{q^{+}}\int d^{2}k_{\bot }\int_{0}^{1}dx 
\frac{\theta (x)\theta (1-x)[-2x+1]}{x(1-x)(q^{-}-\frac{m_{1\bot }^{2}}{
q^{+}x}-\frac{m_{3\bot }^{2}}{q^{+}(1-x)})}\times  \notag \\
&&\frac{1}{K^{-}-\frac{m_{2\bot }^{2}}{q^{+}x+K_{i}^{+}}-\frac{m_{3\bot
}^{2} }{q^{+}(1-x)}}.
\end{eqnarray}

The component $J_{II}^{+}$ of the electromagnetic current in this interval
is linearly proportional $q^{+}$ so that in the limit $q^{+}\rightarrow 0$,
this integral vanishes. This shows that the $``+"$ component of the
electromagnetic current for the bound state of two bosons do not have pair
production contribution. The same happens with the transverse component of
the current, that is, in the limit $q\rightarrow 0$ 
\begin{equation}
J^{+,\bot }=J_{I}^{+,\bot }.
\end{equation}

We see that for the $J^{-}$ component we have to sum two terms in order to
maitain covariance in the light front formalism. When we use Breit's
reference frame, with $q_{z}=0$, we must have $J^{+}=J^{-}$. This equality
comes from a rotation by an $180^0$ angle around $x$-axis, with $q_{x}\neq 0$%
. This rotation transformation in the light front formalism is of non
kinematical nature and when applied to the component $J^+$ of the
electromagnetic current creates pairs. When we include the pair creation in
the light front formalism, we have the correct parity transformation and
rotation. The parity operator for the transformation from $z$ to $-z$ is non
kinematical, so we expect that the property associated with parity that
comes from this operation be destroyed if the pair creation term is not
considered.

\section{Conclusion}

We have demonstrated that the propagator of two bosons in a background
field, has a non-vanishing contribution coming from the pair creation by the
photon. In particular, in an example of bound state with constant vertex, we
demonstrated that the $J^-$ current component in the Breit's reference frame
($q^{+}=0$) has a non-zero contribution from the process of pair creation by
the photon. This conclusion is reached as long as we first have $q^{+}$
different from zero, integrating in $k^{-}$ and then taking the limit $%
q^{+}\rightarrow 0$. The integration in $k^{-}$ and the limit $%
q^{+}\rightarrow 0$ does not commute in general.

In the process of these calculations it has been pointed out that the
emergence of a non-vanishing contribution from pair production by the
interacting photon is naturally achieved by extending the region of allowed
quanta solutions in the light-front, that is, extending the Fock space of
positive quanta to include relevant solutions from the Fock space of
negative quanta. It also means that the myth of light front trivial vacuum
must be forever abandoned.

It has been demonstrated that the inclusion of the pair production term in
the light front formalism is of capital importance for the validity of
rotational symmetry for the electromagnetic current of a bound state of two
bosons in the model of a constant vertex \cite{9}. In the case of components 
$J^{+}$ and $\overrightarrow{J}_{\perp }$ we concluded that the pair
creation term does not contribute in the limit $q^{+}\rightarrow 0$. For the 
$J^{-}$ component, however, we have shown that we must take into account the
pair production so that rotational symmetry be satisfied in the limit $%
q^{+}\rightarrow 0$.

\section*{Acknowledgments}

J.H.O. Sales thanks the hospitality of the Institute for Theoretical
Physics, UNESP, where part of this work has been performed.

\section{Appendix : Current for two free bosons}

To describe the electromagnetic current for a system composed of two free
bosons, we study the process in which two bosons of the same mass $m$
propagate forward in time and in a given instant in the light front $\bar{x}%
^{+}$ one of them interacts with an electromagnetic field. In the following
we calculate the components of two non interacting boson current in an
external electromagnetic field, when total momenta before and after the
absorption of photon being $K_{i}^{+}>0$ and $K^{+}>0$ respectively.

The Lagrangean density that involves scalar field and electromagnetic field
in interaction is given by 
\begin{equation}
\pounds =D_{\mu }\phi D^{\mu }\phi ^{\ast }-m^{2}\phi ^{\ast }\phi .
\label{3.0.0}
\end{equation}

The derivative between the scalar field and the electromagnetic field is
contained in the covariant derivative $D_{\mu}\phi $.

In the calculation of the propagator for a particle in a background field we
use the interaction Lagrangean of a scalar field and electromagnetic field.
As we have already mentioned, the interaction between the scalar and
electromagnetic field is contained in the first term of (\ref{3.0.0}), so
that the interaction Lagrangean is 
\begin{equation}
\pounds _{\text{I}}=ieA^{\mu }\left( \phi \partial _{\mu }\phi ^{\ast }-\phi
^{\ast }\partial _{\mu }\phi \right) +e^{2}A^{\mu }A_{\mu }\phi \phi ^{\ast
}.  \label{3.1.0}
\end{equation}

The Lagrangean (\ref{3.1.0}) shows immediately that there are two types of
vertices. The first term corresponds to a vertex containing a photon and two
scalar particles. The second vertex contains two photons and two scalar
particles.

Using the concept of generating functional $Z\left[ J\right] $, or
vacuum-vacuum transition amplitude in the presence of a external source $%
J\left( x\right) $, we write 
\begin{eqnarray}
Z\left[ J\right] &=&\int \mathcal{D}\phi \exp \left\{ i\int d^{4}x\left[ 
\pounds \left( \phi \right) +J\left( x\right) +\frac{i\varepsilon }{2}\phi %
\right] \right\} \varpropto  \label{3.1.1} \\
&&\left\langle 0,\infty \right. \left\vert 0,-\infty \right\rangle ^{J} 
\notag
\end{eqnarray}
where $\pounds =\pounds _{\text{0}}+\pounds _{\text{I}}$ and 
\begin{equation*}
\pounds _{\text{0}}=\partial _{\mu }\phi \partial ^{\mu }\phi ^{\ast
}-m^{2}\phi ^{\ast }\phi
\end{equation*}

The Green functions are the expectation values of the time ordered product
of field operators in vacuum and according to (\ref{1.3.4}) can be written
in terms of functional derivatives of the generating functional $Z_{0}\left[
J\right] $. That is: 
\begin{equation}
G\left( x_{1},...,x_{n}\right) =\left\langle 0\right\vert T\left( \phi
\left( x_{1}\right) ...\phi \left( x_{n}\right) \right) \left\vert
0\right\rangle  \label{3.2.1}
\end{equation}
which are the $n$-point Green functions of the theory, where 
\begin{equation}
\left\langle 0\right\vert T\left( \phi \left( x_{1}\right) ...\phi \left(
x_{n}\right) \right) \left\vert 0\right\rangle =\frac{1}{i^{n}}\left. \frac{
\delta ^{n}Z_{0}\left[ J\right] }{\delta J\left( x_{1}\right) ...\delta
J\left( x_{n}\right) }\right\vert _{J=0}.  \label{3.2.3}
\end{equation}

Green functions for field theories are extremely important because they are
intimately related to the matrix elements of the scattering matrix $S$ from
which we can calculate quantities measured directly in the experiments such
as scattering processes where the cross section for a given reaction is
measured, decay of a particle in two or more where we can measure the mean
life of particles involved, etc..

The propagator is associated to the Green function equation as: 
\begin{equation}
G(t-t^{\prime })=-iS(t-t^{\prime }).  \label{gpro}
\end{equation}

The Green function or the propagator describes completely the evoltuion for
the quantum system. In this present case we are using the propagator for
``future times''. We could also have defined the propagator ``backwards'' in
time.

The propagation of a free particle with spin zero in four dimensional
space-time is represented by the covariant Feynman propagator 
\begin{equation}
S(x^{\mu })=\int \frac{d^{4}k}{\left( 2\pi \right) ^{4}}\frac{ie^{-ik^{\mu
}x_{\mu }}}{k^{2}-m^{2}+i\varepsilon },  \label{cov}
\end{equation}%
where the coordinate $x^{0}$ represents the time and $k^{0}$ the energy. We
are going to calculate this propagation in the light front, that is, for
times $x^{+}$.

We make the projection of the propagator for a boson in time associated to
the null plane rewriting the coordinates in terms of time coordinate $x^{+}$
and the position coordinates $(x^{-}$ and $\vec{x}_{\perp })$. With these,
the momenta are given by $k^{-}$, $k^{+}$ and $\vec{k}_{\perp }$, and
therefore we have 
\begin{equation}
S(x^{+})=\frac{1}{2}\int \frac{dk_{1}^{-}}{\left( 2\pi \right) }\frac{ie^{ 
\frac{-i}{2}k_{1}^{-}x^{+}}}{k_{1}^{+}\left( k_{1}^{-}-\frac{k_{1\perp
}^{2}+m^{2}}{k_{1}^{+}}+\frac{i\varepsilon }{k_{1}^{+}}\right) }.
\label{lf1}
\end{equation}

The Jacobian of the transformation $k^{0}$, $\vec{k}\rightarrow k^{-},k^{+},%
\vec{k}_{\perp }$ is equal to $\frac{1}{2}$ and $k^{+}$, $k_{\bot }$ are
momentum operators.

Evaluating the Fourier transform, we obtain 
\begin{equation}
\widetilde{S}(k^{-})=\int dx^{+}e^{\frac{i}{2}k^{-}x^{+}}S(x^{+}),
\label{transf}
\end{equation}
where we have used 
\begin{equation}
\delta (\frac{k^{-}-k_{1}^{-}}{2})=\frac{1}{2\pi }\int dx^{+}e^{\frac{i}{2}
\left( k^{-}-k_{1}^{-}\right) x^{+}},  \label{delta}
\end{equation}
and the property of Dirac's delta \textquotedblleft
function\textquotedblright\ 
\begin{equation}
\delta \left( ax\right) =\frac{1}{a}\delta \left( x\right) ,  \label{prop}
\end{equation}
and we get 
\begin{equation}
\widetilde{S}(k^{-})=\frac{i}{k^{+}\left( k^{-}-\frac{k_{\perp }^{2}+m^{2}}{
k^{+}}+\frac{i\varepsilon }{k^{+}}\right) },  \label{prop1}
\end{equation}
which describes the propagation of a particle forward to the future and of
an antiparticle backwards to the past. This can be oberved by the
denominator which hints us that for $x^{+}>0$ and $k^{+}>0$ we have the
particle propagating forward in time of the null plane. On the other hand,
for $x^{+}<0$ and $k^{+}<0$ we have an antiparticle propagating backwards in
time.

\end{document}